\documentclass[runningheads]{svmult}
\usepackage{rotating}   
\usepackage{makeidx}   
\usepackage{graphicx}  
\usepackage{subeqnar}  
\usepackage{multicol}  
\usepackage{cropmark} 
\usepackage{physprbb}  
\newcommand{\ap}{$\bar{\mbox{p}}$}
\newcommand{\Hi}{\mbox{H}}
\newcommand{\W}{\mbox{W}}
\newcommand{\Z}{\mbox{Z}}
\newcommand{\X}{\mbox{X}}
\newcommand{\e}{\mbox{e}}
\newcommand{\qqbar}{\mbox{q}\bar{\mbox{q}}}
\newcommand{\ppbar}{\mbox{p}\bar{\mbox{p}}}
\newcommand{\bbbar}{\mbox{b}\bar{\mbox{b}}}
\newcommand{\stat}{\mbox{stat}}
\newcommand{\sys}{\mbox{sys}}
\newcommand{\lumi}{\mbox{lumi}}

\begin{document}
\title*{Recent D\O\ Results at Run II}
%
%
%
%
%
\author{Stefan S\"oldner-Rembold (on behalf of the D\O\ 
collaboration)\thanks{submitted to the proceedings of
the 4th International Conference on Physics Beyond the Standard Model,
Beyond the Desert '03, Schloss Ringberg, Tegernsee, Germany, 9-14 June 2003
}}
%
%
%
\institute{Department of Physics and Astronomy,
University of Manchester, Manchester, M13 9PL, United Kingdom}

\maketitle              

\begin{abstract}
The D\O\ experiment is taking data with 
an upgraded detector since March 2001. 
The integrated luminosity taken in Run~II has now
exceeded that taken in Run~I. 
Selected physics results obtained with this data set are
presented.
\end{abstract}

\section{Introduction}
The D\O\ experiment is designed to study high energy proton-antiproton
interactions at Fermilab's Tevatron collider in Batavia close to 
Chicago (USA). Protons (p) and anti-protons (\ap) collide at a
centre-of-mass energy $\sqrt{s}=1.96$~TeV. 
A wide variety of physics topics can be studied at the Tevatron.
Measurement of top-quark properties, electroweak precision variables
and B-hadron properties including CP-violation and mixing, as well
as searches for the Higgs boson, Supersymmetry (SUSY) and other 
phenomena beyond the Standard Model. 

\begin{figure}[htbp] 
\begin{center} 
\includegraphics[width=0.65\textwidth]{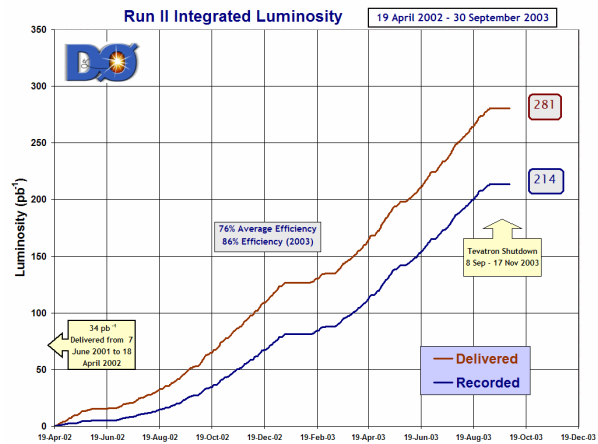} 
\end{center} 
\caption[]{
Run~II integrated luminosity for the period April 2002 - September 2003.}
\label{fig1} 
\end{figure} 

The Run~I data taking period lasted from 1992 to
1996 and produced many important results, among them the discovery of
the top quark. 
Both the Tevatron and the detectors have been upgraded
significantly in the following years. 
Compared to Run~I the Tevatron's centre-of-mass energy
$\sqrt{s}$ has increased from 1.8~TeV to 1.96~TeV. 
The integrated luminosity at the end of Run~II is expected to be a
factor 50 higher than the Run~I integrated luminosity of about 
125~pb$^{-1}$. 

The D\O\ detector has also been upgraded significantly: 
Around the interaction region a new tracking system has
been installed consisting of the Silicon Microvertex Tracker (SMT)
and the Central Fibre Tracker (CFT). The CFT consists of eight  
axial layers of scintillating fibre aligned along the beam axis
and eight layers of stereo fibres. The tracking detector
is located within a 2 Tesla Solenoid. Further improvements
include new preshower detectors surrounding the solenoid, the
Forward Muon Detector, the Forward Proton Detector, and 
improved front-end electronics, trigger and data acquisition.
The Liquid Argon Calorimeter, the Central Muon Detector and 
the muon toroid, which is used to provide a separate momentum
measurement for the muons, are retained from Run~I.
\begin{figure}[htbp] 
\begin{turn}{-90}
\includegraphics[width=0.75\textwidth]{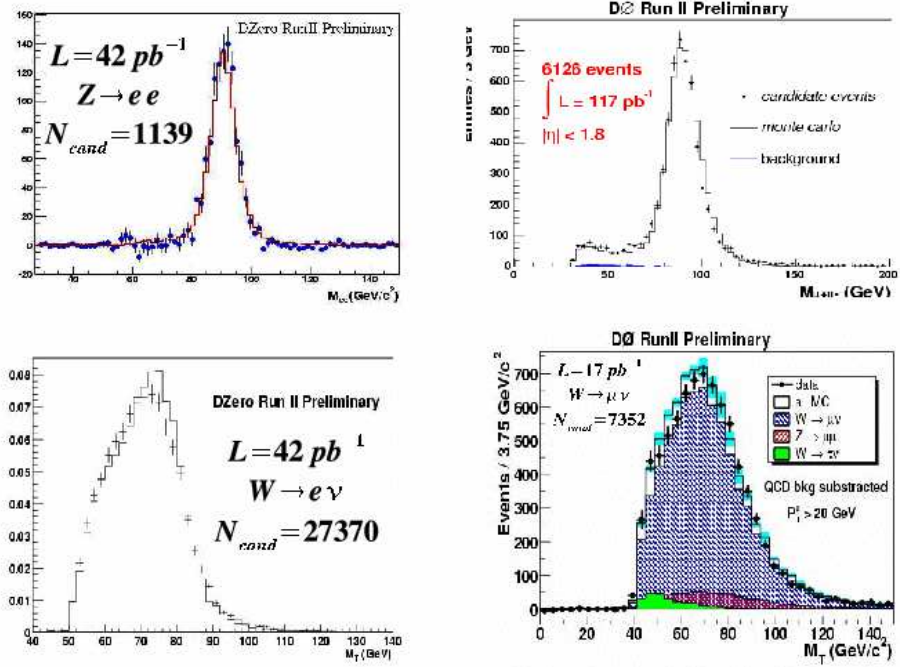} 
\end{turn}
\caption[]{
top: invariant di-lepton mass for $\Z\to \e\e$ and
$\Z\to \mu\mu$ candidate event;
bottom: transverse mass for $\W\to \e\nu$ and $\W\to \mu\nu$
candidate events. 
}
\label{fig2} 
\end{figure} 

In this report a first look at the Run~II data taken since
April 2001 is presented. 
The Run~II collected integrated luminosity is shown in Fig.~1. 
First a measurement of W and Z production is reported
since it is both a fundamental physics measurement and a 
benchmark for the performance of the detector.
Due to the topic of this conference the remaining presentation concentrates
on searches for new physics beyond the 
Standard Model (Large Extra Dimensions, Leptoquarks and SUSY) and
on the search for the Higgs Boson. 
In cases where the results have been updated since the conference,
the updated results are being shown.

\section{Production of W and Z Bosons}
The measurement of the production of W and Z bosons  
is a central part of the D\O\ physics programme. 
The long-term goal is to perform precision measurements 
of electroweak variables
like the W mass and width, to study gauge couplings and to
constrain the parton densities of the proton. 

These processes are also important benchmarks for understanding
the resolution and efficiency of the detector.
The W and Z production cross-sections times branching ratios (BR)
for the processes $ \ppbar \to \W + \X \to \ell \nu + \X$ and 
$\ppbar \to \Z + \X \to \ell \ell + \X$ are measured for
electron and muon final states ($\ell=\e,\mu$).
In the final Run~II data sample we expect to observe more
than 10$^5$ Z decays and more than 10$^{6}$ W decays
into electrons or muons.  
Since the theoretical cross-sections for W and Z production
are calculable with high precision in the Standard Model,
these events can also provide alternative measurements of
the integrated luminosity with small experimental and theoretical errors.

The selection criteria for the $\Z \to  \mu^+\mu^-$ events
require two oppositely charged muons with a transverse momentum
greater than 15 GeV within the pseudorapidity region 
$|\eta|<1.8$ where $\eta=-\ln\tan{\theta/2}$.
The di-muon invariant mass $m_{\mu\mu}$ must be greater than
30~GeV. Further cuts on the energy 
of tracks and calorimeter clusters in a halo around the muon
direction are applied to ensure that the muons are isolated.
The background contribution is estimated to be $(0.6\pm0.3)\%$.
arising from $\bbbar$ production in QCD events and from cosmic rays
and $(0.5\pm0.1)\%$ due to $\Z\to\tau^+\tau^-$ events.
 
\begin{table}
\caption{
Preliminary measurements of the cross-section times branching ratios
for the different W and Z final states. The uncertainty due
to the luminosity measurement is currently $10\%$.}
\begin{center}
\renewcommand{\arraystretch}{1.4}
\setlength\tabcolsep{5pt}
\begin{tabular}{|c|c|c|}
\hline\noalign{\smallskip}
Process & $\sigma\cdot$BR  & L \\
\noalign{\smallskip}
\hline
\noalign{\smallskip}
$\ppbar\to \Z + \X \to \mu\mu + \X$ & $261.8 \pm 5.0 (\stat) \pm 8.9 (\sys) \pm 26.2 (\lumi)$~pb & 117~pb$^{-1}$ \\ \hline 
$\ppbar\to \Z + \X \to \e\e + \X$   & $294 \pm 11 (\stat) \pm 8 (\sys) \pm 29 (\lumi)$~pb & 41.6~pb$^{-1}$ \\ \hline 
$\ppbar\to \W + \X \to \mu\nu + \X$ & $3226 \pm 128 (\stat) \pm 100 (\sys) \pm 323 (\lumi)$~pb & 17.3~pb$^{-1}$ \\ \hline 
$\ppbar\to \W + \X \to \e\nu + \X$  & $3054 \pm 100 (\stat) \pm 86 (\sys) \pm 305 (\lumi)$~pb & 41.6~pb$^{-1}$ \\ \hline 
\hline
\end{tabular}
\end{center}
\label{Tab1.1a}
\end{table}

Events with $\W\to\mu\nu$ decays are selected by requiring
exactly one muon with a transverse momentum greater than 20~GeV.
The muon must be isolated and additional cuts against muons
from cosmic rays are applied. The missing transverse energy
in the event, calculated from the calorimeter information and corrected
for the muon transverse momentum, must
exceed 20~GeV. About $17\%$ of the selected events are expected
to be background, mainly from $\Z\to\mu\mu$, $\W\to\tau\nu$ and
QCD events.

\begin{figure}[htbp] 
\begin{center} 
\includegraphics[width=0.7\textwidth]{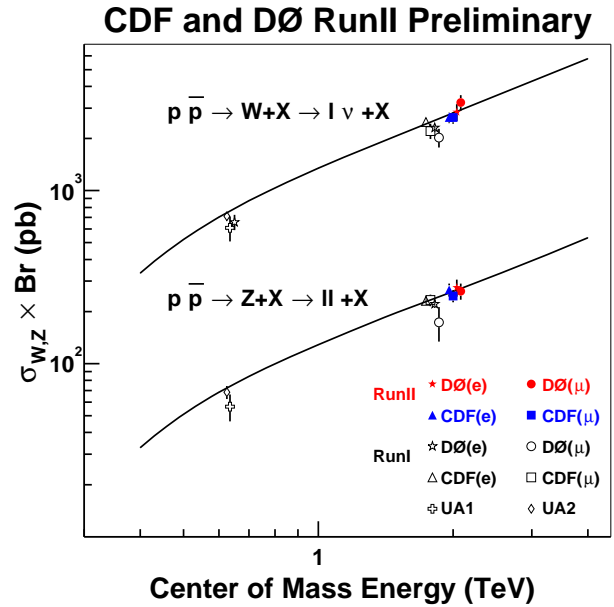} 
\end{center} 
\caption[]{
Summary plot for the W and Z production cross-section
in electron and muon final states as a function of $\sqrt{s}$.
}
\label{fig3} 
\end{figure} 

The identification of electrons in $\W \to \e\nu$  and $\Z \to \e\e$  
events is based on requiring one or two electrons with a transverse momentum of
more than 25 GeV in the central calorimeter 
$(|\eta|<1.1)$. For the $\W \to \e\nu$ channel the missing transverse
energy due to the neutrino must be larger than 25 GeV.

Figure~2 shows the di-lepton invariant mass for the Z candidates and 
the transverse mass for the W candidates. The cross-sections are given 
Table~1. In Figure 3, the measured cross-sections are compared with
the cross-sections measured by the CDF collaboration and with
the cross-sections measured at lower $\sqrt{s}$. Within the uncertainties
good agreement is observed with the evolution of the cross-section
expected from a NLO calculation~\cite{nlo}.

\section{Searches for New Physics}
The D\O\ experiment has started to search for rare event topologies
which are not consistent with the expectations of the Standard Model.
Most searches are based on leptons (mainly
electrons and muons) and on missing transverse energy due to neutrinos.
Similar topologies can be interpreted in many different models which
predict new physics beyond the Standard Model. With the current integrated
luminosity some of the searches start to become competitive with
similar searches performed at Run~I, at LEP and at HERA. In this report
only a few highlights can be presented.

\subsection{Large Extra Dimensions}
It has been suggested in the context of string theory
that the apparent 
Planck scale, $M_{\rm Pl}\simeq 10^{19}$~GeV, which is the scale
where gravity becomes comparable in strength to other interactions, 
can be related to physics at the TeV scale and is therefore
accessible to experimental tests at current high energy experiments.
Arkani, Dimopoulos, and Dvali~\cite{add} have proposed a framework
where the hierarchy problem is solved by assuming that the
fundamental Planck scale $M_{\rm S}$ is 
similar in size to the electroweak scale. 
\begin{figure}[htbp] 
\begin{center} 
\includegraphics[width=0.9\textwidth]{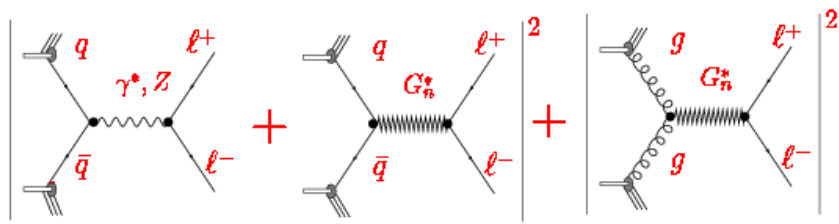} 
\end{center} 
\caption[]{
Diagrams for di-lepton production in the presence of Large
Extra Dimensions~\protect\cite{gupta}.}
\label{fig-gupta}
\end{figure} 
This is accomplished by relating the apparent Planck scale $M_{\rm Pl}$ 
to the Planck scale in a $(3+n)$-dimensional theory,
\begin{equation}
M_{\rm Pl} = (M_{\rm PL}^{(3+n)})^{n+2} R^{n}= M_{\rm S}^{n+2} R^{n},
\end{equation}
where $n$ is the number of Extra Dimensions and $R$ is
the compactification radius.
Assuming for the fundamental Planck scale $M_{\rm S} \simeq 1$~TeV, 
$n=2$ corresponds to $R\simeq 1$~mm,
which is accessible to gravity experiments. 
Only accelerator experiments, however, can study the region $n>2$.

\begin{figure}[htbp] 
\begin{center} 
\includegraphics[width=0.7\textwidth]{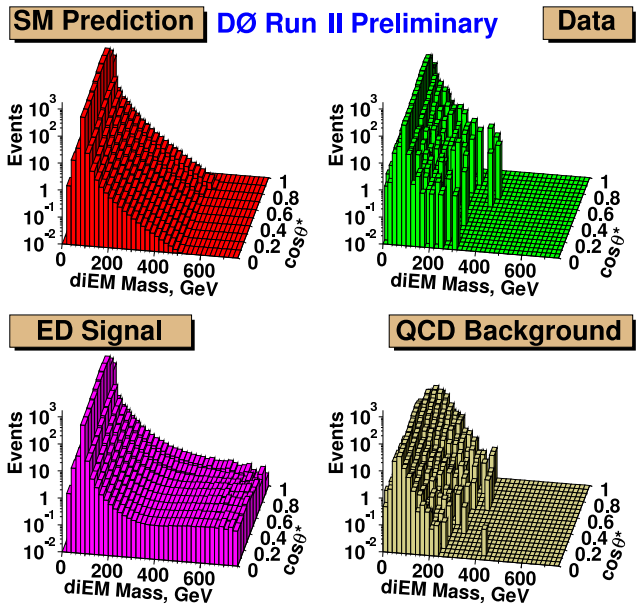} 
\end{center} 
\caption[]{
Two-dimensional $(M, \cos\theta^*)$ distributions for the ee
and $\gamma\gamma$ final states. The expected Standard Model (SM) background,
the expected signal due to Large Extra Dimensions (ED) for $\eta_G=2$,
the QCD background and the data are shown separately.
}
\label{fig-ed} 
\end{figure} 
 
Large Extra Dimensions are expected to manifest themselves
through the effects of virtual graviton exchange in di-fermion or
di-boson production (see Fig.~\ref{fig-gupta}).
They are parametrised via a single
variable $\eta_{G}={\cal F}/M_{\rm S}^4$, where $\cal F$ is
a dimensionless parameter of order unity. Different formalisms
are used to define $\cal F$.

The cross-section is parametrised by
\begin{equation}
\frac{{\rm d}^2 \sigma}{{\rm d}M {\rm d} \cos \theta*} = 
f_{\rm SM} + f_{\rm interference}\eta_{G}+f_{\rm Kaluza-Klein}\eta_{G}^2,
\end{equation}
where $f$ are functions of the di-fermion (or di-boson) invariant mass $M$
and the scattering angle $\cos\theta^*$ in the centre-of-mass frame.
The first term is due to the Standard Model, the last
term due to Kaluza-Klein graviton exchange and the second term
due to interference. Graviton exchange would modify the
$(M,\cos\theta^*)$ distributions.

A search for Large Extra Dimensions in the 
di-electron and di-photon channels has been performed
using $L=130$~pb$^{-1}$ of data. A similar analysis has been
performed in the di-muon channel using $100$~pb$^{-1}$ of data.
For the di-electron and di-photon analysis two electromagnetic objects
in the calorimeter, which are consistent with an electron or
photon hypothesis and have
transverse energies greater than 25~GeV, are required for
the final selection. The two-dimensional distribution 
of the invariant mass M and the scattering angle $\cos\theta^*$
is shown in Fig.~\ref{fig-ed}.

New limits on the fundamental Planck scale $M_{\rm S}$ of 1.28 TeV 
(in the GRW convention~\cite{bib-grw}) are derived from a fit
to the $(M,\cos\theta^*)$ distribution, exceeding limits obtained 
with similar statistics in Run~I~\cite{bib-edrun1}. 
Combined with the published Run~I result, this corresponds to a lower limit 
of $M_{\rm S} > 1.37$~ TeV, 
which is the most stringent limit on Large Extra Dimensions to date.

\subsection{Leptoquarks}
Leptoquarks (LQ) are coloured spin 0 or spin 1 particles carrying both
baryon (B) and lepton (L) quantum numbers. They appear in many extensions
of the Standard Model as a consequence of the symmetry between the lepton
and quark sectors. The Buchm\"uller-R\"uckl-Wyler (BRW) model~\cite{bib-brw}
assumes lepton and baryon number conservation.
Moreover the simplifying assumption is made that a given leptoquark couples to 
just one family of fermions.
Leptoquarks may decay into either a charged lepton
and a quark or into a neutrino and a quark. 
The branching ratio of the decay into a charged lepton
and a quark is commonly denoted by $\beta$.
\begin{figure}[htbp] 
\begin{center} 
\includegraphics[width=0.42\textwidth]{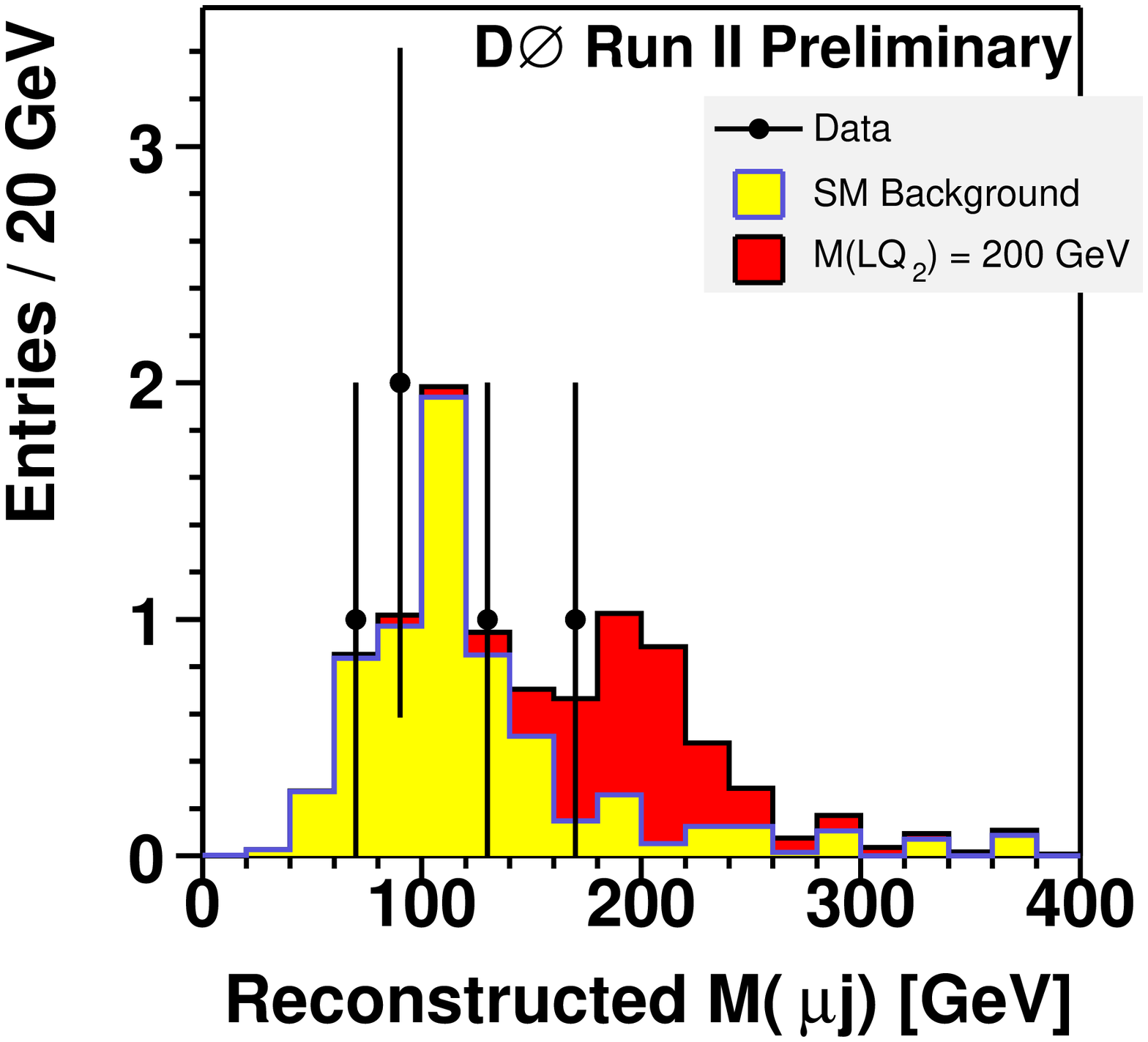}
\includegraphics[width=0.55\textwidth]{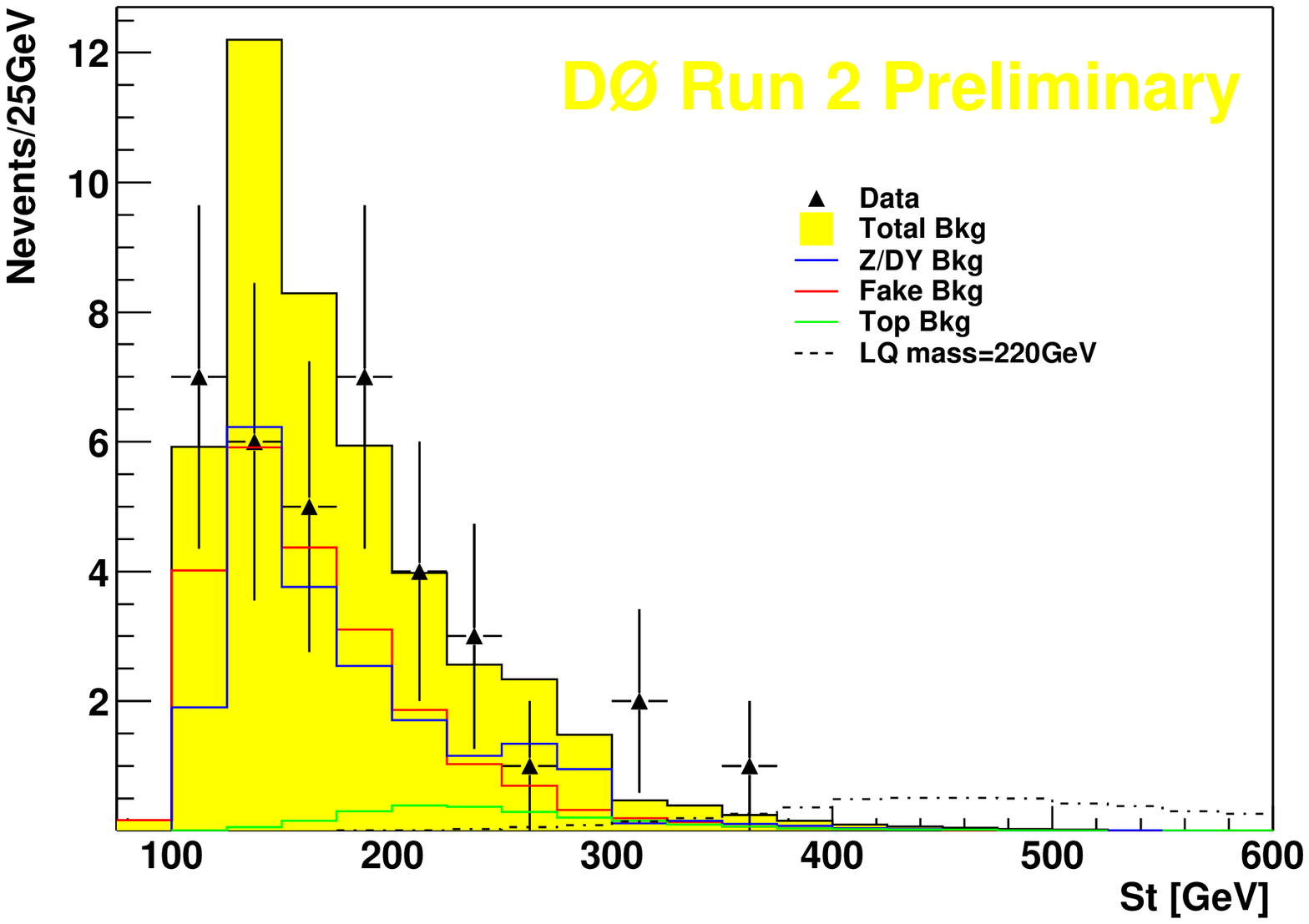}
\end{center} 
\caption[]{
left plot: Reconstructed $\mu$+jet mass for all 
di-muon+di-jet events with $M(\mu\mu) > 110$~GeV. 
There are two possibilities to combine the two highest-$p_{\rm T}$
muons with the two highest transverse energy jets. Only the
combination with the smaller mass difference of the two leptoquark 
candidates of the event is chosen,
and the reconstructed LQ mass of this event is the mean of the 
masses of the two  $\mu$+jet systems; 
right plot: $S_{T}$ distribution of di-electron + di-jet 
events in data (triangles) compared to background (histogram) 
after applying a veto cut against Z events. 
The dashed histogram is the $S_T$ distribution for a 220 GeV LQ signal.
}
\label{fig-lq} 
\end{figure} 

The D\O\ experiment has searched for scalar (spin 0) first generation
and second generation leptoquarks decaying into eq and $\mu$q
final states, respectively. 
The scalar leptoquarks are pair produced.
The mass limits are therefore independent of the Yukawa coupling
$\lambda$. Assuming $\beta=1$ the final states are eeqq for the
first generation and $\mu\mu$qq for the second generation
leptoquark. Both analyses therefore require at least two
jets with a transverse energy of more than 25~GeV in addition
to two charged leptons with high transverse momentum $p_{\rm T}$.

The reconstructed $\mu$+jet mass distribution is shown in Fig.~\ref{fig-lq}
together with a simulated second generation LQ signal with a mass of 200~GeV.
The scalar sum $S_T$ of the transverse energy of the two electrons
and the two leading jets in the eeqq channel is found to
give a good separation of signal and background. The $S_T$
distribution is shown in Fig.~\ref{fig-lq} together with
a simulated first generation LQ signal with a mass of 210~GeV.
No evidence of LQ production is observed in any of the channels,

A data sample of 135~pb$^{-1}$ is used for the eeqq channel.
A lower mass limit of 231 GeV for a first generation
scalar leptoquark at $95\%$ Confidence 
Level (CL) is obtained, in the case of $\beta=1$. 
The combined mass limit for a  
scalar first generation leptoquark is 253 GeV from 
D\O\ Run~II and Run~I results, which is the most stringent limit to date.

The integrated luminosity used for the $\mu\mu$qq channel 
is $104$~pb$^{-1}$. 
No excess of the data over the expected background is found, and 
the $95\%$ CL mass limit for a scalar second
generation leptoquarks assuming $\beta= 1$ is found to be $186$~GeV. 
 
In addition, the case where one of the two leptoquarks 
decays into an electron and a quark and the other in a neutrino and a 
quark is also considered. 
Assuming $\beta=0.5$, an upper mass limit of 156 GeV
at $95\%$ CL is obtained for a data set with an integrated
luminosity of $121$~pb$^{-1}$.

\subsection{Gauge Mediated Supersymmetry (GMSB)}
D\O\ is searching for supersymmetric particles in a variety
of different theoretical frameworks. A particularly interesting
and clear signature is provided by 
Gauge Mediated Supersymmetry Breaking (GMSB). 
In GMSB models, the gravitino ($\tilde{G}$) 
is the lightest supersymmetric particle (LSP).

\begin{figure}[htbp] 
\begin{center} 
\includegraphics[width=0.7\textwidth]{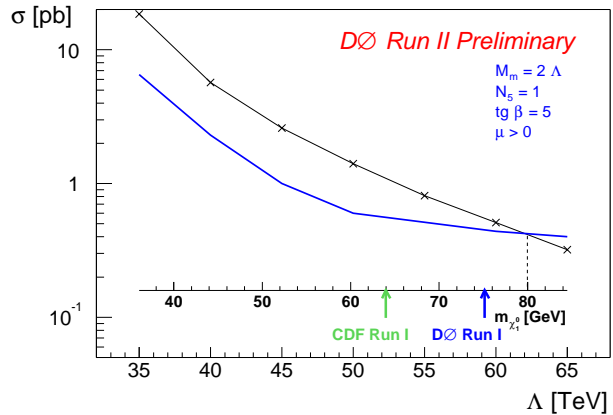} 
\end{center} 
\caption[]{
GMSB cross-section as a function of 
the SUSY breaking scale $\Lambda$ for $\tan\beta=5$, where 
$\tan\beta$ is the
ratio of the Higgs vacuum expectation values. 
The $95\%$ confidence levels limits and the Run~I CDF and D\O\ results
are also shown~\protect\cite{run1}.
}

\label{fig-gmsb} 
\end{figure} 

The neutralino is the next-to-lightest supersymmetric particle (NLSP)
in the scenario considered here. A typical process within this scenario is
$$\ppbar \to  \tilde{\chi}_1^{0}\tilde{\chi}_1^{0} + \X
\to \gamma\gamma+ \tilde{G}\tilde{G}+X,$$
where the gravitino $\tilde{G}$ remains undetected.

An inclusive search for signals of new physics is therefore carried out 
in the di-photon plus missing energy channel.
The analysis requires two electromagnetic clusters in the central
part of the calorimeter  
with a transverse energy greater than 20~GeV
and a missing transverse energy greater than 35 GeV. 
In total $1.38 \pm 0.30$ events are expected from the 
Standard Model background and 
zero events are observed. 
The GMSB cross-section as a function of the SUSY
breaking scale $\Lambda$ is shown in Fig.~\ref{fig-gmsb}.
In the absence of a signal,
lower bounds of $m_{{\tilde{\chi}}^{0}_{1}}>80$~GeV and 
$m_{{\tilde{\chi}}^{\pm}_{1}}>144$~GeV are derived for the neutralino
and chargino mass, respectively.

\section{Searches for the Higgs Boson}
\begin{figure}[htbp] 
\begin{center} 
\includegraphics[width=0.45\textwidth]{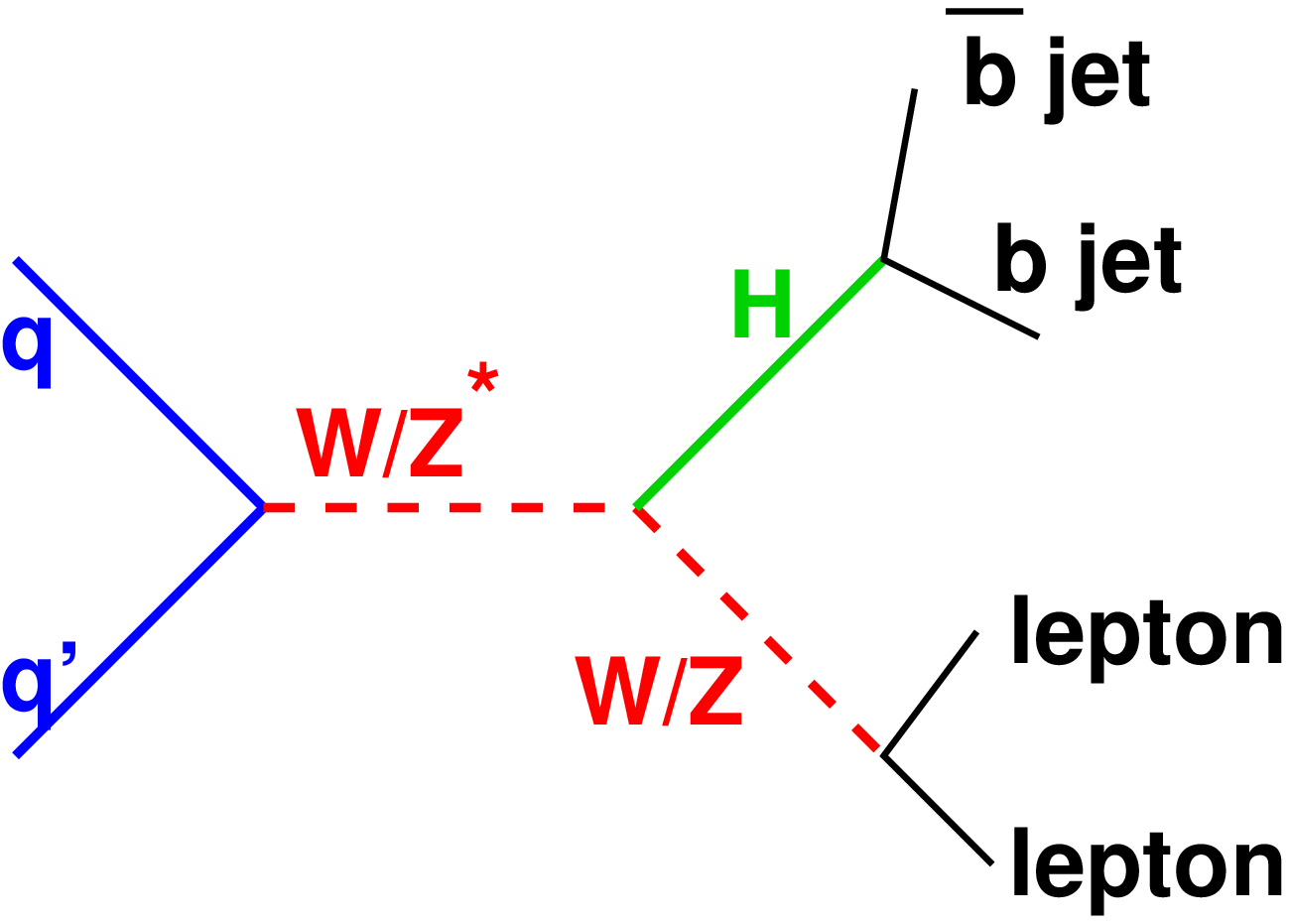} 
\end{center} 
\caption[]{
Higgs production in the channel
$\mbox{qq}'\to  \W^* \to \W \Hi$.}
\label{fig-higgsw} 
\end{figure}

The search for the Higgs boson is one of the most
ambitious physics goals of the Tevatron.  
Production of the Standard Model (SM) Higgs boson through the
Higgs-strahlung processes, 
$\mbox{qq}'\to  \W^* \to \W \Hi$ and $\qqbar \to \Z^* \to \Z \Hi$, 
where the W and Z boson decay leptonically, is the most promising search
channel at the Tevatron (Fig.~\ref{fig-higgsw}).
The SM cross-section times branching ratio for producing
a Higgs
with a mass of $120$~GeV is about 30~fb for the WH process 
if the W bosons decays into electron or muons.
The signal rate in the current data
sample corresponding to an integrated luminosity of about $200$~pb$^{-1}$
is therefore much too small to be observed. 

\begin{figure}[htbp] 
\begin{center} 
\includegraphics[width=0.45\textwidth]{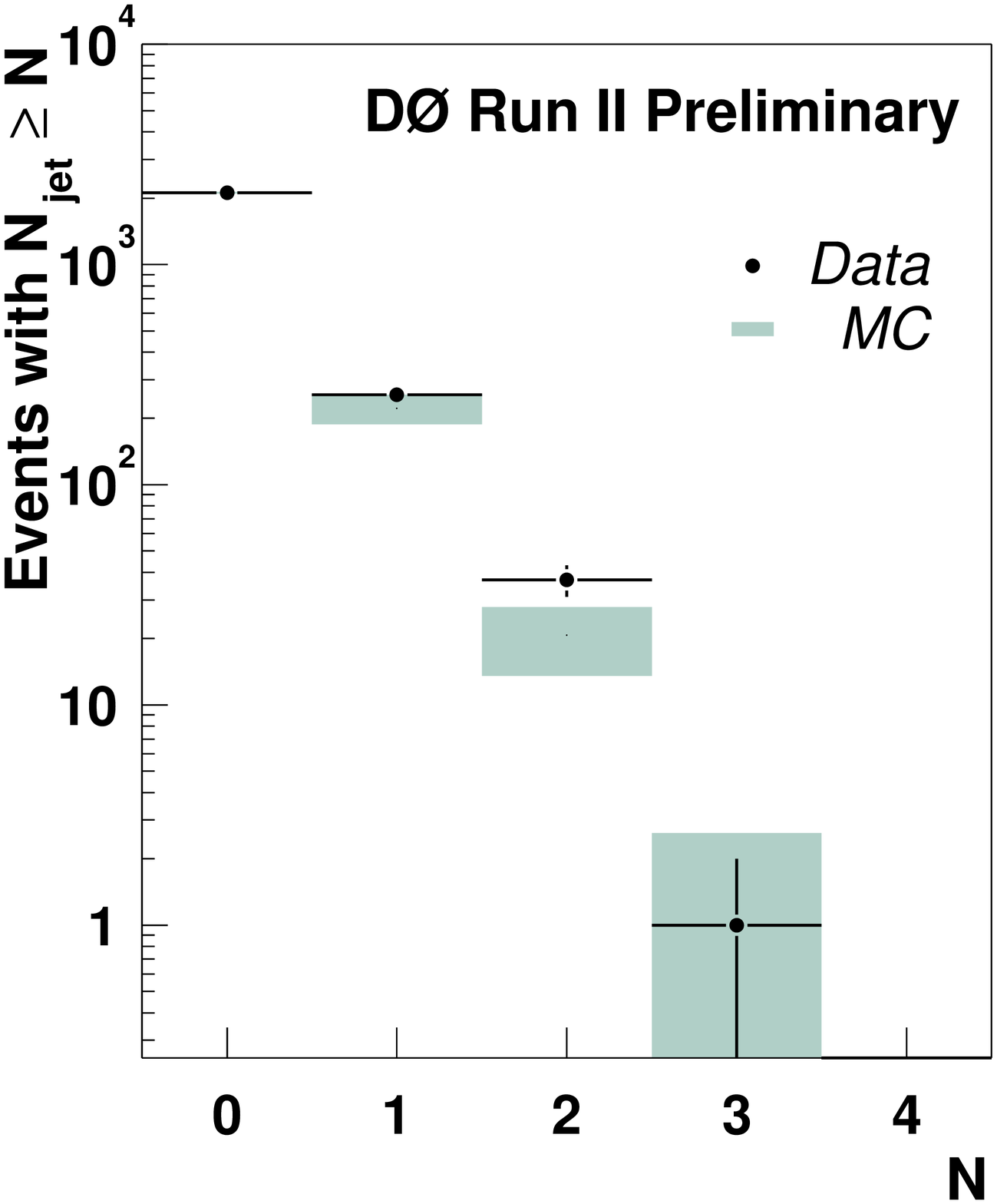} 
\includegraphics[width=0.45\textwidth]{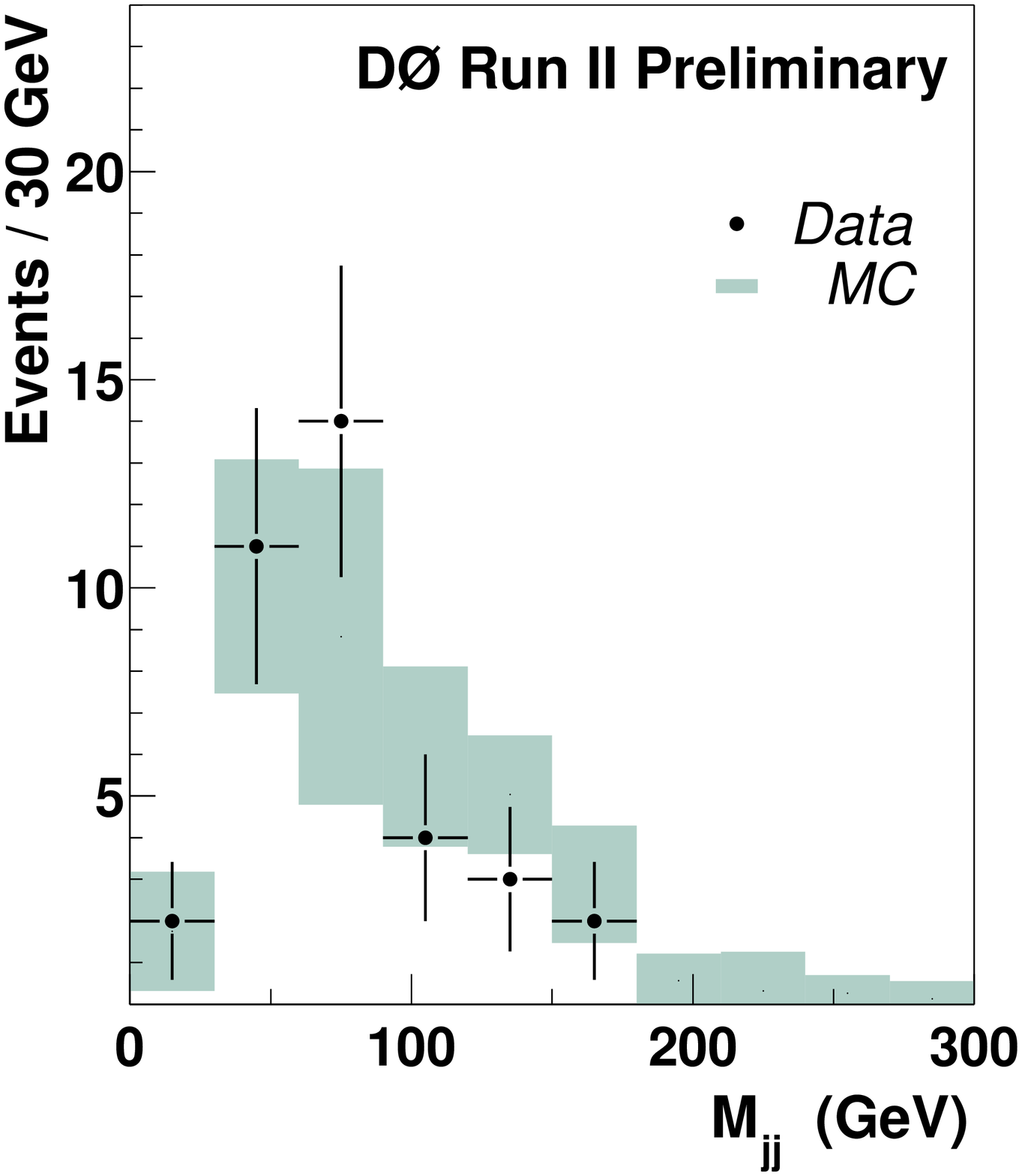} 
\end{center} 
\caption[]{
The data points show the combined data 
for Z events with e+jets and $\mu$+jets final states.
The shaded bands show the Monte Carlo prediction. 
The width of the band reflects the (currently statistics limited) 
range allowed by the jet energy calibration.
left plot: inclusive jet multiplicities;
right plot: di-jet invariant mass.}
\label{fig-higgsjets} 
\end{figure} 

The ''Road to the Higgs Discovery'', however, is well defined.
Searching for the SM Higgs boson in the WH (or ZH) channel
first requires a precise understanding of 
the production of W and Z bosons.
These results have been presented in Section 2. 

The next step is related to understanding \mbox{(di-)jet} production in
events with W and Z production. 
In Fig.~\ref{fig-higgsjets}
the measured jet multiplicity and di-jet invariant mass 
in $\Z\to\mu\mu$ and $\Z\to\e\e$ events
are compared
to a simulation using the PYTHIA Monte Carlo program.
The jet selection requires the jet to have a transverse
momentum larger than 20~GeV and it has to lie within
the pseudorapidity range $|\eta|<2.5$.
The band used for the Monte Carlo prediction shows
the uncertainty due to the calorimeter energy scale.
Reasonably good agreement is observed between data and Monte
Carlo.

For SM Higgs mass below 135~GeV, the Higgs boson
primarily decays into $\bbbar$ final states.
Jets originating from b-quarks can be identified using
the lifetime $\tau\simeq 1.5$~ps of B hadrons which translates
into a decay length of the order 1~mm in the detector.
The tagging of b-jets is performed by measuring the impact parameter of tracks,
the reconstruction of secondary vertices and by identifying
semi-leptonic b decays. 
An important tool is the
Silicon Microvertex Detector (SMT) which has been shown
to measure the impact parameter of tracks with large
transverse momentum  with a resolution of $\simeq 15 \mu$m.

\begin{figure}[htbp]
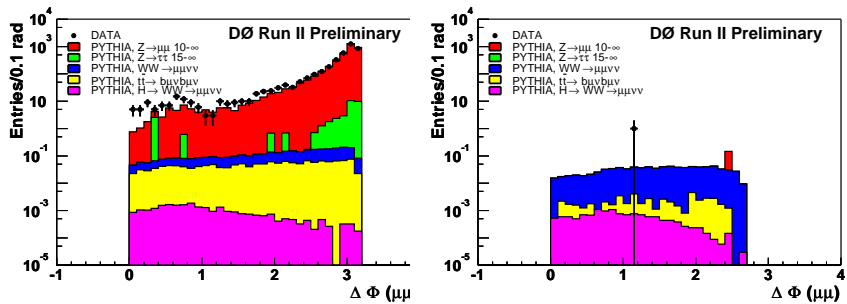
 
\begin{center} 
\includegraphics[width=0.45\textwidth]{dphi_mumu_data_mc_new_1.epsi} 
\includegraphics[width=0.45\textwidth]{dphi_mumu_data_mc_new_4.epsi} \\
\end{center} 
\caption[]{
Azimuthal angle between the two
muons, $\Delta\phi(\mu\mu)$ after a loose pre-selection
and after the final cuts. 
}
\label{fig-azi} 
\end{figure} 

If the mass of the Higgs particle is in the range of 135 to 200 GeV, it
will predominantly decay to W pairs. A search for Higgs decays
in W pairs has been performed for decays into ee, e$\mu$ and $\mu\mu$
final states using data corresponding to an integrated luminosity
of 118 pb$^{-1}$. The electrons have to be within a
pseudorapidity range of $|\eta|<1.3$ and the muons within
$|\eta|<2$. The cuts on the transverse momenta vary between
7~GeV and 20~GeV depending on the final state and the Higgs mass studied.
For the $\mu\mu$ channel the azimuthal angle between the two
muons, $\Delta\phi(\mu\mu)$, is shown in Fig.~\ref{fig-azi}.
Since the spins of the two W bosons from the Higgs decay
are correlated,  $\Delta\phi(\mu\mu)$ is on average smaller
for signal than for background events.

\begin{figure}[htbp] 
\begin{center} 
\includegraphics[width=0.45\textwidth]{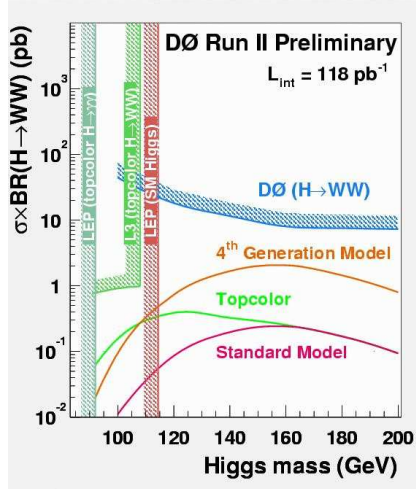} 
\end{center} 
\caption[]{
Higgs production cross section times $\mbox{BR}(H\to WW)$ 
excluded with the current D\O\ data
(and LEP experiments), along with the expectations from various models.
This result is a combination of the ee, e$\mu$ and $\mu\mu$ analyses.
}
\label{fig-WW} 
\end{figure}

After the final event selection using a sequence of cuts
to suppress background one event is observed in 
good agreement with the
Monte Carlo expectation for the sum of all SM processes of 
$0.95 \pm 0.12$ events.
Since no signal is observed in this channel and in the e$\mu$ and
ee channels, a combined limit on the Higgs production cross section times 
the branching ratio $\mbox{BR}(H\to WW)$ at $95\%$ CL is obtained
in Fig.~\ref{fig-WW}. A SM Higgs boson in the WW decay
channel cannot be excluded yet with the current integrated
luminosity, but the sensitivity is within less than a factor
of ten for a 4th generation model.

In addition to the results presented here
the D\O\ experiment is searching for Higgs production
in several other scenarios beyond the Standard Model. 
Results have recently been presented on Higgs decays into two
photons and on doubly-charged Higgs bosons decaying into two muons.

\section{Summary}
First results of the D\O\ experiment on W and Z production,
searches for Large Extra Dimensions, Leptoquarks
and Gauge Mediated Supersymmetry have been reported.
With the recorded integrated
luminosity many search results are already improving on
previous limits obtained at Run~I and LEP.
The Tevatron is currently the only running collider
at the high energy frontier and many exciting results
are expected in future years.

\section*{Acknowledgement}
I would like to thank Hans Volker Klapdor-Kleingrothaus and
Irina Krivosheina for the excellent organisation of the scientific
and social aspects of the conference. The wide
variety of topics presented at the conference and the
beautiful setting of Schloss Ringberg made the
conference a most enjoyable experience. 
I also want to thank my colleagues at D\O\ for their
help in preparing this report and Emily Nurse for
her careful reading of the manuscript.

%

\end{document}